\documentclass[preprint,1p,times]{elsarticle}
\usepackage[cmex10]{amsmath}
\usepackage{amssymb}
\usepackage{amsmath}
\usepackage{mathtools}
\usepackage{accents}
\usepackage{epsfig}
\usepackage{epstopdf}
\usepackage{mdwmath}
\usepackage{mdwtab}
\usepackage{stfloats}
\fnbelowfloat
\usepackage{IEEEtrantools}
\usepackage{color}

\biboptions{sort&compress}

\journal{Signal Processing}

\newcommand{\ACal}{\mathcal{A}}
\newcommand{\Ab}{\mathbf{A}}
\newcommand{\Bb}{\mathbf{B}}
\newcommand{\bb}{\mathbf{b}}

\newcommand{\eb}{\mathbf{e}}
\newcommand{\Ub}{\mathbf{U}}
\newcommand{\Vb}{\mathbf{V}}
\newcommand{\Wb}{\mathbf{W}}
\newcommand{\Xb}{\mathbf{X}}

\newcommand{\Xbs}{\mathbf{X}^*}

\newcommand{\xb}{\mathbf{x}}
\newcommand{\xbs}{\xb^*}
\newcommand{\xbd}{\xb^{\downarrow}}
\newcommand{\yb}{\mathbf{y}}

\newcommand{\Zerb}{\mathbf{0}}

\newcommand{\Rbb}{\mathbb{R}}
\newcommand{\DimDef}{\Rbb^{n_1 \times n_2}}
\newcommand{\LOp}{\ACal: \Rbb^{n_1 \times n_2} \rightarrow \Rbb^m}
\newcommand{\nDef}{n \triangleq \min(n_1,n_2)}

\newcommand{\Sigb}{\boldsymbol{\sigma}}

\newcommand{\SigbWb}{\Sigb(\Wb)}

\newcommand{\sumi}{\sum_{i \geq 2}}
\newcommand{\dtt}{\delta_{2t}}

\newcommand{\vb}{\mathbf{v}}

\newcommand{\NullDef}{\mathcal{N}(\ACal)}

\DeclareMathOperator{\rank}{rank}

\DeclareMathOperator{\diag}{diag}

\DeclareMathOperator{\nullS}{null}

\def\MinEdt{black}
\def\ROneEdt{black}

\begin{document}

\begin{frontmatter}



\title{Performance Guarantees for Schatten-$p$ Quasi-Norm Minimization in Recovery of Low-Rank Matrices\tnoteref{t1}}
\tnotetext[t1]{This work was supported in part by Iran National Science Foundation under contract 91004600. The work of the first author was supported in part by a travel scholarship from Ericsson Research during his visit at the Communication Theory Lab., KTH- Royal Institute of Technology.}


\author[add1]{Mohammadreza~Malek-Mohammadi\corref{cor1}}
\ead{m.rezamm@ieee.org}
\author[add1]{Massoud~Babaie-Zadeh}
\ead{mbzadeh@yahoo.com}
\author[add2]{Mikael Skoglund}
\ead{skoglund@ee.kth.se}

\cortext[cor1]{Corresponding author. Tel: +98 912 5123401.}

\address[add1]{Electrical Engineering Department, Sharif University of Technology, Tehran 1458889694}
\address[add2]{Communication Theory Lab, KTH- Royal Institute of Technology, Stockholm, 10044, Sweden}

\begin{abstract}
 We address some theoretical guarantees for Schatten-$p$ quasi-norm minimization ($p \in (0,1]$) in recovering low-rank matrices from compressed linear measurements. Firstly, using null space properties of the {\color{\MinEdt}measurement} operator, we provide a sufficient condition for exact recovery of low-rank matrices. This condition guarantees unique recovery of matrices of ranks equal or larger than what is guaranteed by nuclear norm minimization. Secondly, this sufficient condition leads to a theorem proving that all restricted isometry property (RIP) based sufficient conditions for $\ell_p$ quasi-norm minimization generalize to Schatten-$p$ quasi-norm minimization. Based on this theorem, we provide a few RIP-based recovery conditions.
\end{abstract}


\end{frontmatter}

% \linenumbers

\section{Introduction}
Matrix rank minimization constrained to a set of underdetermined linear equations, known as affine rank minimization (ARM), has numerous applications in signal processing and control theory \cite{RechFP10,CandP10}. An important special case of this optimization problem is \emph{Matrix Completion} (MC) in which one aims to recover a matrix from partially observed entries \cite{CandP10}. Applications of ARM and MC include collaborative filtering \cite{CandP10}, machine learning \cite{AmitFSU07}, quantum state tomography \cite{GrosLFBE10}, ultrasonic tomography \cite{ParhKOV13}, spectrum sensing \cite{KoocMBS15}, direction-of-arrival estimation \cite{MaleJOKB14}, and RADAR \cite{KaloP14}, among others.

Rank minimization under affine equality constraints is generally formulated as
\begin{equation}
\min_{\Xb}\rank(\Xb) \quad \text{subject to} \quad \ACal(\Xb)=\bb, \label{RM}
\end{equation}
where $\Xb \in \DimDef$, $\LOp$ is a given linear operator (measurement operator), and $\bb \in \Rbb^m$ is the vector of measurements. In case of incomplete measurements, $m$ is less than $n_1n_2$, or, usually, $m \ll n_1n_2$. Problem \eqref{RM} is generally NP-hard \cite{RechFP10}, yet there are many efficient algorithms to solve relaxed or approximated versions of it. Nuclear norm minimization (NNM), proposed in \cite{RechFP10}, replaces the rank with its tightest convex relaxation which leads to
\begin{equation}
\min_{\Xb} \|\Xb\|_* \quad \text{subject to} \quad \ACal(\Xb)=\bb, \label{NNM}
\end{equation}
where $\|\Xb\|_* \triangleq \sum\nolimits_{i=1}^{r}{\sigma_i(\Xb)}$ denotes matrix nuclear norm in which $\sigma_i(\Xb)$ is the $i$th largest singular value %\footnote{In this paper, we always assume that singular values of matrices are sorted in descending order.}
of $\Xb$ and $r$ is the rank of the matrix $\Xb$. It has been proven that, under some sufficient conditions, \eqref{RM} and \eqref{NNM} share the same unique solution; see, e.g., \cite{RechXH11,CandP10}.

The nuclear norm of a matrix is equal to the $\ell_1$ norm of a vector formed by the singular values of the same matrix. Consequently, inspired by experimental observations and theoretical guarantees showing superiority of $\ell_p$ quasi-norm minimization to $\ell_1$ minimization in Compressive Sampling (CS) \cite{Bara07}, another approach in \cite{MohaF12,MarjS12} replaces the rank function with the Schatten-$p$ quasi-norm resulting in \begin{equation} \label{pNM}
\min_{\Xb} \|\Xb\|_p^p \quad \text{subject to} \quad \ACal(\Xb)=\bb,
\end{equation}
where $\|\Xb\|_p \triangleq \big( \sum\nolimits_{i=1}^{r} \sigma_i^p(\Xb) \big)^{1/p}$ for some $p \in (0,1)$ denotes the Schatten-$p$ quasi-norm. While the above problem is nonconvex, it is observed that numerically efficient implementations of \eqref{pNM} outperforms NNM \cite{MohaF12,MarjS12,MajuW11}.

In practice, there is often some noise in measurements, so measurement model is updated to $\ACal(\Xb)+\eb=\bb$, where $\eb$ is the vector of measurement noise. To robustly recover a minimum-rank solution,
equality constraints are relaxed to $\|\ACal(\Xb)-\bb\|_2 \leq \epsilon$, where $\|\cdot\|_2$ denotes the $\ell_2$ norm of a vector and $\epsilon \geq \|\eb\|_2$
 is some constant \cite{CandP10}. Therefore, \eqref{pNM} is modified to
\begin{equation} \label{pNM_Noisy}
\min_{\Xb} \|\Xb\|_p^p \quad \text{subject to} \quad \|\ACal(\Xb)-\bb\|_2 \leq \epsilon.
\end{equation}

Though there are several theoretical studies concerning $\ell_p$ quasi-norm minimization in the CS literature (see, for example, \cite{GribN07,FoucL09,CharS08,WangXT11}), only a few papers deal with performance guarantees of Schatten-$p$ quasi-norm minimization ($p$SNM). In \cite{OymaMFH11}, authors propose a necessary and sufficient condition for exact recovery of low-rank matrices using null space properties of $\ACal$. However, the sufficient condition is not sharp and seems to be stronger than that of NNM. In contrast, it is well known that finding the global solution of $\ell_p$ quasi-norm minimization in CS scenario is superior to $\ell_1$ minimization \cite{GribN07,FoucL09,CharS08}. Therefore, when one considers the strong parallels between CS and ARM (see \cite{RechFP10} for a comprehensive discussion) and superior experimental performance of $p$SNM in comparison to NNM, he/she expects weaker recovery conditions. We will show that this intuition is indeed the case by providing a sharp sufficient condition, and proving that, using \eqref{pNM}, one can uniquely find matrices with equal or larger ranks than those of recoverable by NNM.

In addition, we further exploit this sufficient condition and extend a result from \cite{OymaMFH11} to prove that all restricted isometry property (RIP) based results for recovery of sparse vectors using $\ell_p$ quasi-norm minimization generalize to Scahtten-$p$ quasi-norm minimization {\color{\ROneEdt}with no change}. In particular, extending some results of \cite{FoucL09}, we will show that if $\delta_{2r} < 0.4531$, then all low-rank or approximately low-rank matrices with at most $r$ dominant singular values can be recovered accurately from noisy measurements via \eqref{pNM_Noisy}. This generalization also proves that, for some sufficiently small $p > 0$, if $\delta_{2r+2} < 1$, then, program \eqref{pNM_Noisy} recovers all matrices with at most $r$ large singular values from noisy measurements accurately. Furthermore, another RIP-based sufficient condition will be presented which is sharper than a threshold in \cite{FoucL09} for small values of $p$.

The rest of this letter is organized as follows. After introducing some notations, in Section \ref{Main}, we will present our performance analysis. Section \ref{Proofs} is devoted to the proofs of the main results which is followed by conclusion.

\emph{Notations}: {\color{\MinEdt}A vector is called $k$-sparse if it has $k$ nonzero components.} $\xbd$ denotes a vector obtained by sorting elements of $\xb$ in terms of magnitude in descending order, and $\xb^{(k)}$ designates a vector consisted of {\color{\MinEdt}the} $k$ largest elements (in magnitude) of $\xb$. Let $\langle \xb,\yb \rangle \triangleq \xb^T\yb$ be the inner product of $\xb$ and $\yb$ and $\|\xb\|_2 \triangleq \langle \xb,\xb \rangle^{\frac{1}{2}}$ stand for the Euclidean-norm. $\ell_p$ quasi-norm of $\xb$ for $p \in (0,1)$ is defined as $\| \xb \|_p \triangleq \big( \sum_{i} x_i^p \big)^{1/p}$, where $x_i$ is the $i$th entry of $\xb$. For any matrix $\Xb \in \DimDef$, define $\nDef$. It is always assumed that singular values of matrices are sorted in descending order, and $\Sigb(\Xb) = (\sigma_1(\Xb),\ldots,\sigma_n(\Xb))^T$ is the vector of singular values of $\Xb$. $\|\Xb\|_F \triangleq  \sqrt{\sum_{i=1}^{n} \sigma_i^2(\Xb)}$ denotes the Frobenius norm. Furthermore, let $\Xb = \Ub \diag(\Sigb(\Xb)) \Vb^T$ denotes the singular value decomposition (SVD) of $\Xb$, where $\Ub \in \Rbb^{n_1 \times n}$ and $\Vb \in \Rbb^{n_2 \times n}$. $\Xb^{(r)} = \Ub \diag(\sigma_1(\Xb),\ldots,\sigma_r(\Xb),0,\cdots,0) \Vb^T$ represents a matrix obtained by keeping the $r$ largest singular values in the SVD of $\Xb$ and setting others to 0. For a linear operator $\LOp$, let $\mathcal{N}(\ACal) \triangleq \{\Xb \in \DimDef : \ACal(\Xb) = \Zerb, \Xb \neq \Zerb\} = \nullS(\ACal)\backslash\{\Zerb\}$. For a set $S$, $|S|$ denotes its cardinality.

\section{Main Results} \label{Main}
\subsection{A null space condition}
%Null space properties have been used as an important tool to characterize the uniqueness of solutions of algorithms in recovery of sparse vectors, see e.g., \cite{DonhE03,GribN07}.
%Null space properties have been used as a tool to characterize the conditions under which the success of sparse vector or low-rank matrix recovery is guaranteed, see e.g., \cite{DonhE03,GribN07}. Particularly, \cite{OymaH10} derives a necessary and sufficient condition for successful reconstruction of minimum-rank solutions via NNM. These results were later extended to $p$SNM in \cite{OymaMFH11}, %However, in the previous work,
In \cite{OymaMFH11}, exploiting null space properties of $\ACal$, a necessary and sufficient condition for successful reconstruction of minimum-rank solutions via \eqref{pNM} are derived, yet there is a gap between these conditions. In this paper, we close this gap by introducing the following lemma, which is mainly based on a result from \cite{ZhanQ10}, and prove that the necessary condition in \cite{OymaMFH11} is also sufficient. Moreover, we will show that, using $p$SNM, one can uniquely recover all matrices with equal or larger rank than those of uniquely recoverable by NNM. %The proofs of the following theorems are left to Section \ref{Proofs1}.

\newtheorem{Lem1}{Lemma}
\begin{Lem1} \label{NSPNec}
All matrices $\Xb \in \DimDef$ of rank at most $r$ can be uniquely recovered by \eqref{pNM}, provided that, $\forall \Wb \in \NullDef$,% if and only if
\begin{equation*}
\sum_{i=1}^{r} \sigma_i^p(\Wb) < \sum_{i=r+1}^{n} \sigma_i^p(\Wb).
\end{equation*}
%where $\NullDef \triangleq \nullS(\ACal)\backslash\{\Zerb\}$ and $n \triangleq \min(n_1,n_2)$.
%The proof is left to Section \ref{appA}.
\end{Lem1}

{\color{\ROneEdt}It is worth mentioning that the sufficient condition in Lemma \ref{NSPNec} is weaker than the corresponding sufficient condition in \cite{OymaMFH11} which, according to our notations, is formulated as
\begin{equation*}
\sum_{i=1}^{2r} \sigma_i^p(\Wb) < \sum_{i=2r+1}^{n} \sigma_i^p(\Wb).
\end{equation*}
Since $\sum_{i=1}^{r} \sigma_i^p(\Wb) \leq \sum_{i=1}^{2r} \sigma_i^p(\Wb)$ and $\sum_{i=r+1}^{n} \sigma_i^p(\Wb) \geq \sum_{i=2r+1}^{n} \sigma_i^p(\Wb)$, the sufficient condition in Lemma \ref{NSPNec} is less restrictive than \cite[Theorem 3]{OymaMFH11}.} Based on the above sufficient condition, %let us define
%\begin{equation*}
%\theta_p(r,\ACal) \triangleq \sup_{\Wb \in \NullDef} \frac{\sum_{i=1}^{r} \sigma_i^p(\Wb)}{\sum_{i=1}^{n} \sigma_i^p(\Wb)}.
%\end{equation*}
%$\theta_p(r,\ACal)$ extends similar parameter defined in \cite{GribN07} for $\ell_0$-norm minimization in CS and describes the uniqueness of \eqref{pNM} as it gives a unique solution if $\theta_p(r,\ACal) < 1/2$.
we have the following proposition which is a routine extension of \cite[Theorem 5]{GribN07}.

\newtheorem{Prop1}{Proposition}
\begin{Prop1} \label{Sup}
Let $r_p^*(\ACal)$ and $r_1^*(\ACal)$ denote the maximum ranks such that all matrices $\Xb$ with $\rank(\Xb) \leq r_p^*(\ACal)$ and $\rank(\Xb) \leq r_1^*(\ACal)$ can be uniquely recovered by \eqref{pNM} and \eqref{NNM}, respectively. Then $r_p^*(\ACal) \geq r_1^*(\ACal)$ for any $p \in (0,1)$.
%$r_p^*(\ACal) \geq r_1^*(\ACal)$ for $p \in (0,1)$.
\end{Prop1}

\subsection{RIP-based conditions}
%A more well known approach to analyze recovery conditions for sparse vectors and low-rank matrices from incomplete measurements is restricted isometry property. %This concept was first used in vector case \cite{Cand08} and later on was generalized to rank minimization in \cite{RechFP10,CandP11}. %Here, the definitions for vectors and matrices are recalled. This definition slightly differs from \cite{RechFP10}, but equals to \cite{CandP11,CaiA12}.
Inspired by the strong parallels between CS and ARM, \cite{OymaMFH11} simplifies generalization of some results on $\ell_1$ norm minimization to nuclear norm minimization. Remarkably, it shows that all RIP-based conditions for stable and robust recovery of sparse vectors through $\ell_1$ norm minimization directly generalize to nuclear norm minimization. %Considering available theoretical background for $\ell_p$ quasi-norm minimization in the CS context,
%Utilizing the above sufficient condition, we generalize one of the results in \cite{OymaMFH11} to show that all RIP based recovery conditions for $\ell_p$ quasi-norm minimization are extended to $p$SNM.
{\color{\ROneEdt}Furthermore, \cite{OymaMFH11} proves a similar equivalence between RIP-based conditions for recovery of sparse vectors via $\ell_p$ quasi-norm minimization and recovery of low-rank matrices using $p$SNM. Nevertheless, the established equivalence in \cite[Lemma 14]{OymaMFH11} is not as strong as one might expect. In essence, it shows an equivalence between RIP conditions for recovery of \emph{$2k$-sparse} vectors and RIP conditions for reconstruction of \emph{rank $k$} matrices. However, it is natural to have the equivalence between sparsity and rank of the same order. Utilizing Lemma 1, we make the order of sparsity and rank equal to $k$ in the aforementioned equivalence.} To that end, first, formulation of $\ell_p$ quasi-norm minimization as well as the definitions of RIP for vector and matrix cases are recalled.

In $\ell_p$ quasi-norm minimization, the program
\begin{equation} \label{lpmin_Noisy}
\min_{\xb} \|\xb\|_p^p \quad \text{subject to} \quad \|\Ab \xb-\bb_v\|_2 \leq \epsilon
\end{equation}
is used to estimate a sparse vector $\xb \in \Rbb^{m_v}$ from noisy measurements $\bb_v = \Ab \xb + \eb_v$ in which $\Ab \in \Rbb^{n_v \times m_v}$ and $\bb_v \in \Rbb^{n_v}$ are known and $\eb_v$ is noise vector with $\| \eb_v \|_2 \leq \epsilon$.

\newtheorem{Def1}{Definition}
\begin{Def1}[\hspace{-0.05em}\cite{Cand08}]
 For matrix $\Ab$ and all integers $k \leq m_v$, the restricted isometry constant (RIC) of order $k$ is the smallest constant $\delta_k(\Ab)$ such that
 \begin{equation*}
 (1-\delta_{k}(\Ab)) \|\xb\|_2^2 \leq \|\Ab \xb\|_2^2 \leq (1+\delta_{k}(\Ab)) \|\xb\|_2^2
 \end{equation*}
 holds for all vectors $\xb$ with sparsity at most $k$.
 \end{Def1}

\newtheorem{Def2}[Def1]{Definition}
\begin{Def2}[\hspace{-0.05em}\cite{OymaMFH11}]
 For linear operator $\ACal$ and all integers $r \leq n$, the RIC of order $r$ is the smallest constant $\delta_r(\ACal)$ such that
 \begin{equation*}
 (1-\delta_{r}(\ACal)) \|\Xb\|_F^2 \leq \|\ACal(\Xb)\|_2^2 \leq (1+\delta_{r}(\ACal)) \|\Xb\|_F^2
 \end{equation*}
 holds for all matrices $\Xb$ with rank at most $r$.%, where $\|\cdot\|_F$ stands for the Frobenius-norm of matrices.
 \end{Def2}

 The following theorem %whose proof is given in Section \ref{Proofs}
 formally shows how the results are extended to $p$SNM.

\newtheorem{Thm1}{Theorem}
\begin{Thm1} \label{GentoMat}
Let $\xb_0 \in \Rbb^{m_v}$ be any arbitrary vector, $\bb_v = \Ab \xb_0 + \eb_v$, and $\xbs$ denote a solution to \eqref{lpmin_Noisy} to recover $\xb_0$. %, where $\bb_v \in \Rbb^{n_v}$ is known and $\eb_v$ is noise with $\|\eb_v\|_2 \leq \epsilon$.
Likewise, let $\Xb_0 \in \DimDef$ be any arbitrary matrix, $\bb = \ACal( \Xb ) + \eb$, and $\Xbs$ denote a solution to \eqref{pNM_Noisy} to recover $\Xb_0$. %, where $\bb \in \Rbb^m$ is known and $\eb$ is noise with $\|\eb\|_2 \leq \epsilon$.
Assume that RIP condition $f(\delta_{k_1}(\Ab), \cdots,\delta_{k_u}(\Ab)) < \delta_0$, for some function $f$, is sufficient to have
\begin{IEEEeqnarray*}{llCl}
& \| \xb_0 - \xbs \|_p & \leq & g_1(\xbd_0, \epsilon), \\
& \| \xb_0 - \xbs \|_2 & \leq & g_2(\xbd_0, \epsilon),
\end{IEEEeqnarray*}
for some functions $g_1$ and $g_2$. Then, under the same RIP condition $f(\delta_{k_1}(\ACal), \cdots,\delta_{k_u}(\ACal)) < \delta_0$, we have
\begin{IEEEeqnarray*}{llCl}
& \| \Xb_0 - \Xbs \|_p & \leq & g_1(\Sigb(\Xb_0), \epsilon), \\
& \| \Xb_0 - \Xbs \|_F & \leq & g_2(\Sigb(\Xb_0), \epsilon).
\end{IEEEeqnarray*}
%where $\NullDef \triangleq \nullS(\ACal)\backslash\{\Zerb\}$ and $n \triangleq \min(n_1,n_2)$.
%The proof is left to Section \ref{appA}.
\end{Thm1}

%Before presenting RIP based conditions, let us define residual of the best $r$-rank approximation of a matrix $\Xb$ in terms of the Schatten-$p$ quasi-norm as
%\begin{equation*} \label{pRes}
%\Delta_{r,p}(\Xb) \triangleq \min_{\rank(\Yb) \leq r} \|\Xb - \Yb\|_p.
%\end{equation*}

One of the best uniform thresholds on $\delta_{2k}$ for finding $k$-sparse vectors using $\ell_p$ quasi-norm minimization is given in \cite{FoucL09}. This threshold works uniformly for any $p \in (0,1]$ and covers exact recovery conditions as well as robust and accurate reconstruction of sparse and nearly-sparse vectors from noisy measurements. %Here, we extend the results in \cite{FoucL09} to low-rank and approximately low-rank matrices by means of the following proposition and corollary.
Theorem \ref{GentoMat} simply generalizes the results in \cite{FoucL09} to low-rank matrix recovery by means of the following proposition and corollary. To have a more organized presentation, we use the inequality $\gamma_{2t} \geq (1 + \delta_{2t}) / (1 - \delta_{2t})$, where $\gamma_{2t}$ is the asymmetric RIC defined in \cite{FoucL09}, to state our results in terms of $\dtt$ (the RIC defined herein).

%\newtheorem{Def2}[Def1]{Definition}
%\begin{Def2}
%For linear operator $\ACal$ and all integers $r \leq n$, the asymmetric restricted isometry constants of order $r$ are the best constants $\alpha_r,\beta_r$ such that
%\begin{equation*} \label{ARIP}
%\alpha_r \|\Xb\|_F \leq \|\ACal(\Xb)\|_2 \leq \beta_r \|\Xb\|_F
%\end{equation*}
%holds for all matrices $\Xb$ with rank at most $r$.
%\end{Def2}
%Our results in terms of asymmetric RIP is based on the ratio of the above constants, specifically $\gamma_r \triangleq \beta_r^2 / \alpha_r^2$. Furthermore,

%The proofs of the following theorems have been left to Section \ref{appB}.

\newtheorem{Prop2}[Prop1]{Proposition}
\begin{Prop2} \label{RIPThm}
Let $\Xb_0 \in \DimDef$ be any arbitrary matrix and $\ACal(\Xb_0) + \eb=\bb$, where $\bb \in \Rbb^m$ is known and $\eb$ is noise with $\|\eb\|_2 \leq \epsilon$. Suppose that $\Xbs$ is a solution to \eqref{pNM_Noisy} to recover $\Xb_0$ for some $p \in (0,1]$. If
%\vspace{-1em}
\begin{equation} \label{NoisyRecCond}
\dtt < \frac{2(\sqrt{2}-1)(t/r)^{\frac{1}{p}-\frac{1}{2}}}{2(\sqrt{2}-1)(t/r)^{\frac{1}{p}-\frac{1}{2}} + 1}
\end{equation}
%\vspace{-1em}
holds for some integer $t \geq r$, then
%\vspace{-1em}
\begin{IEEEeqnarray*}{llCl}
& \|\Xb_0-\Xbs\|_p & \leq &  C_1  \|\Xb_0 - \Xb_0^{(r)} \|_p + D_1 r^{\frac{1}{p} - \frac{1}{2}} \epsilon, \\
& \|\Xb_0-\Xbs\|_F & \leq &  C_2 t^{\frac{1}{2} - \frac{1}{p}} \|\Xb_0 - \Xb_0^{(r)} \|_p + D_2 \epsilon.
\end{IEEEeqnarray*}
The constants $C_1, C_2, D_1, D_2$ depend only on $p$$,\dtt$$,t/r$ and are given in \cite[Theorem 3.1]{FoucL09}. In particular, when $\epsilon = 0$ and $\rank(\Xb_0) \leq r$, \eqref{NoisyRecCond} implies that $\Xb_0$ is a unique solution to \eqref{pNM}.
\end{Prop2}

Two important special cases of the above sufficient condition are summarized in the following corollary.

\newtheorem{Cor1}{Corollary}
\begin{Cor1}
The sufficient condition of Proposition \ref{RIPThm} implies the following sufficient conditions too:
\begin{itemize}
  \item $\delta_{2r} < 0.4531$ for any $p \in (0,1]$,
  \item knowing $r$ and $\delta_{2r+2} < 1$, it is possible to find some $p_0$ such that inequality \eqref{NoisyRecCond} holds for all $0 < p < p_0$.
\end{itemize}
\end{Cor1}

%Several RIP based sufficient conditions have been proposed in the literature for accurate and robust recovery of low-rank matrices using NNM. For instance, $\delta_{4r} < \sqrt{2} - 1 \approx 0.4142$ in \cite{CandP11}, $\delta_{3r} < 2\sqrt{5} - 4 \approx 0.4721$, $\delta_{4r} < (8 - \sqrt{40})/3 \approx 0.558$, and $\delta_{5r} < (12 - \sqrt{60})/7 \approx 0.607$ in \cite{MohaF10}, and $\delta_{2r} < 1/2$ and $\delta_r < 1/3$ in \cite{CaiA12}. %Therefore, comparing thresholds obtainable from Theorem \ref{RIPThm} to previous ones, we can summarize that Theorem \ref{RIPThm}
%\begin{enumerate}
%  \item provides the first RIP based sufficient condition for robust and accurate recovery of low-rank matrices using $p$SNM,
%  \item improves the best $\delta_{2r}$ uniform recovery threshold for sparse vector recovery, which is up to our best knowledge $\delta_{2r} < 2(3-\sqrt{2})/7 \approx 0.4531$ in \cite{FoucL09}, to $\delta_{2r} < 1/2$,
%  \item gives a uniform threshold which is as strong as the best result for NNM,
%  \item introduces a recovery condition, $\delta_{2r + 2} < 1$ for any $p \in (0,p_0]$ with some sufficiently small $p_0$, which is close to $\delta_{2r} < 1$, a sufficient condition for the uniqueness of the original problem \eqref{RM} \cite{RechFP10}.
%\end{enumerate}
%In the $p$SNM side, up
%To our best knowledge, Theorem \ref{RIPThm} is the first result that considers reconstruction of approximately low-rank matrices via $p$SNM. Nevertheless,
Theorem \ref{GentoMat} also generalizes other recent RIP-based conditions in $\ell_p$ quasi-norm minimization (e.g., the conditions in \cite{WuC13,HsiaS2013}). %However, to our best knowledge, \cite[Theorem 3.1]{FoucL09} i  %Schatten-$p$ quasi-norm minimization. Still, there is a gap between $\delta_{2r} < 0.4531$ and $\delta_{2r} < 1/2$. Hence, we think that, with a more sophisticated approach like \cite{MoL11,CaiA12}, one may obtain stronger thresholds and improve $\delta_{2r} < 0.4531$. Also, note that $\delta_{2r + 2} < 1$ is close to $\delta_{2r} < 1$ which is a sufficient condition for the %uniqueness of the original problem \eqref{RM} \cite{RechFP10}.
%In addition to the above conditions, Theorem \ref{RIPExactThm2} introduces another sufficient condition for exact recovery, and Theorem \ref{RIPThm2} shows that this sufficient condition also guarantees robust and accurate reconstruction of low-rank matrices.% using $p$SNM.
In addition to the above conditions, below, we introduce another sufficient condition which guarantees robust and accurate reconstruction of low-rank matrices.% using $p$SNM.

\newtheorem{Thm2}[Thm1]{Theorem}
\begin{Thm2} \label{RIPThm2}
Under assumptions of Proposition \ref{RIPThm}, if
\begin{equation} \label{NoisyRecCond2}
\dtt < \frac{(t/r)^{\frac{2}{p}-1} - 1}{(t/r)^{\frac{2}{p}-1} + 1}
\end{equation}
holds for some integer $t \geq r$, then
\begin{IEEEeqnarray*}{rCl}
& \|\Xb_0-\Xbs\|_p \leq &  C_1'  \|\Xb_0 - \Xb_0^{(r)} \|_p + D_1' r^{\frac{1}{p}-\frac{1}{2}} \epsilon, \\
& \|\Xb_0-\Xbs\|_F \leq &  C_2' t^{\frac{1}{2} - \frac{1}{p}} \|\Xb_0 - \Xb_0^{(r)} \|_p + D_2'  \epsilon,
\end{IEEEeqnarray*}
where the constants $C_1', C_2', D_1', D_2'$ depend only on $p,\dtt,t/r$. In particular, when $\epsilon = 0$ and $\rank(\Xb_0) \leq r$, \eqref{NoisyRecCond2} implies that $\Xb_0$ is a unique solution to \eqref{pNM}.
\end{Thm2}

\begin{figure}[h]
\centering
\includegraphics[width=0.42\textwidth]{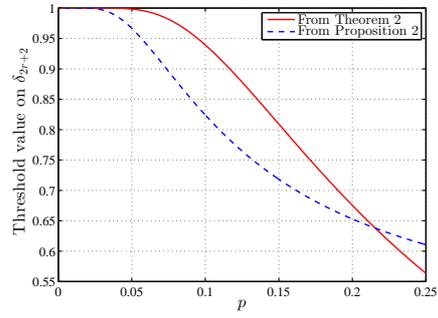}
\vspace{-0.4cm}
\caption{Recovery thresholds from Proposition \ref{RIPThm} and Theorem \ref{RIPThm2} as a function of $p$. $r$ is fixed to 5, and thresholds are independent of matrix dimensions.} \label{fig:pCurve}
\end{figure}

%\medskip
Despite the fact that a uniform recovery threshold cannot be obtained from Theorem \ref{RIPThm2}, substituting $t$ with $r+1$ in \eqref{NoisyRecCond2}, we get
\begin{equation} \label{Smallp2}
\delta_{2r+2} < \frac{(1 + 1/r)^{\frac{2}{p}-1}-1}{(1 + 1/r)^{\frac{2}{p}-1}+1}.
\end{equation}
%Neglecting the constant terms, since $(1 + 1/r)$ in \eqref{Smallp2} grows twice that of \eqref{Smallp1} with decrease of $p$, it is expected that $\delta_{2r+2} < 1$ guarantees accurate and robust recovery for two times larger $p_0$'s than those of Theorem \ref{RIPThm}. Figure \ref{fig:pCurve} shows $\delta_{2r+2}$ thresholds from Theorem \ref{RIPThm} and \ref{RIPThm2} for $r = 5$. As it is clear, Theorem \ref{RIPThm2} threshold becomes sharper than that of Theorem \ref{RIPThm} after passing $p \approx 0.26$. Furthermore, it reaches to 1 for $p \leq 0.04$, while the one from Theorem \ref{RIPThm} approaches to 1 for $p \leq 0.02$.
Fixing $r$ and $\delta_{2r+2}$, let $p_0$ denote the maximum value {\color{\MinEdt}such that} all $p \in (0,p_0)$ satisfy \eqref{NoisyRecCond} for $t = r+1$. Respectively, let $p_0'$ denote the maximum value {\color{\MinEdt}such that} all $p \in (0,p_0')$ satisfy \eqref{Smallp2}. Neglecting the constant terms, since, with the decrease of $p$, the power of $(1 + 1/r)$ in \eqref{Smallp2} grows twice that of in \eqref{NoisyRecCond}, it is expected that \eqref{Smallp2} guarantees accurate recovery for $p_0' \approx \sqrt{p_0}$ when thresholds in the right-hand side of \eqref{NoisyRecCond} and \eqref{Smallp2} tend to 1. Figure \ref{fig:pCurve} shows $\delta_{2r+2}$ thresholds derived from Proposition \ref{RIPThm} and Theorem \ref{RIPThm2} as a function of $p$ for $r = 5$. As it is clear, the threshold given in Theorem \ref{RIPThm2} becomes sharper than that of given in Proposition \ref{RIPThm} after passing $p \approx 0.22$. Furthermore, it reaches to 1 at $p \approx 0.05$, while the one from Proposition \ref{RIPThm} approaches to 1 at $p \approx 0.025$. {\color{\ROneEdt} Recall that $\delta_{2r} < 1$ is a sufficient condition for the success of the original rank minimization problem in \eqref{RM} \cite{RechFP10}. Consequently, the above result shows that, for a larger range of $p$'s, $p$SNM is almost optimal since $\delta_{2r+2}<1$ guarantees its success.}

%\begin{figure}[tb]
%\centering
%\includegraphics[width=0.42\textwidth]{ThCurve.eps}
%\vspace{-0.4cm}
%\caption{Recovery thresholds from Proposition \ref{RIPThm} and Theorem \ref{RIPThm2} as a function of $p$. $r$ is fixed to 5, and thresholds are independent of matrix dimensions.} \label{fig:pCurve}
%\end{figure}

\section{Proofs of results} \label{Proofs}
%\subsection{Proof of Theorems \ref{NSPNec}, \ref{NSPSuf}, and \ref{Sup}} \label{appA}
\subsection{Preliminaries}
We begin with a definition and a few lemmas. %This lemma is originally from \cite{ZhanQ10}.
% Also, recall the following definition.

\newtheorem{Def3}[Def1]{Definition}
\begin{Def3}[\hspace{-0.05em}\cite{HornJ90}]
A function $\Phi(\xb) : \Rbb^n \rightarrow \Rbb$ is called symmetric gauge if it is a norm on $\Rbb^n$ and invariant under arbitrary permutations and sign changes of $\xb$ elements.
\end{Def3}

\newtheorem{Lem3}[Lem1]{Lemma}
\begin{Lem3}[{{\hspace{-0.05em}\cite[Corollary 2.3]{ZhanQ10}}}] \label{SubAddLem}
Let $\Phi$ be a symmetric gauge function and $f: [0,\infty) \rightarrow [0,\infty)$ be a concave function with $f(0)=0$. %, and $\Sigb(\cdot)$ be the vector of SVs.
Then for $\Ab,\Bb \in \DimDef$,
\begin{equation*}
\Phi \Big( f\big(\Sigb(\Ab)\big) - f\big(\Sigb(\Bb)\big) \Big) \leq \Phi \Big( f \big( \Sigb(\Ab - \Bb) \big) \Big),
\end{equation*}
where $f(\xb) = (f(x_1), \ldots, f(x_n))^T$.% and $\Sigb(\Xb) = (\sigma_1(\Xb),\ldots,\sigma_n(\Xb))^T$ denotes vector of singular values of $\Xb$.
\end{Lem3}

\newtheorem{Lem4}[Lem1]{Lemma}
\begin{Lem4} \label{SubAddCor}
Let $\Ab,\Bb \in \DimDef$. For any $p \in (0,1]$,% we have
%\begin{align} \label{mainInEq}
%\sum_{i=1}^{n} \sigma_i^p(\Ab - \Bb) & \geq \sum_{i=1}^{n} |\sigma_i^p(\Ab) - \sigma_i^p(\Bb)| \\
%& \geq \sum_{i=1}^{n} \sigma_i^p(\Ab) - \sigma_i^p(\Bb) \nonumber.
%\end{align}
\begin{equation} \label{mainInEq}
\sum_{i=1}^{n} \sigma_i^p(\Ab - \Bb) \geq \sum_{i=1}^{n} |\sigma_i^p(\Ab) - \sigma_i^p(\Bb)|.
\end{equation}
\begin{IEEEproof}
It is obvious that $\Phi(\xb) = \sum_{i=1}^n |x_i|$ and $f(x) = x^p, p \in (0,1),$ satisfy conditions of Lemma \ref{SubAddLem}. Thus, \eqref{mainInEq} is an immediate result for $p \in (0,1)$. Moreover, \eqref{mainInEq} holds for $p = 1$ \cite{HornJ90}.% of Lemma \ref{SubAddLem}.
\end{IEEEproof}
\end{Lem4}

\newtheorem{Lem5}[Lem1]{Lemma}
\begin{Lem5} \label{InEqLem}
Let $\Wb = \Ub \diag (\SigbWb) \Vb^T$ denote the SVD of $\Wb$. If for some $\Xb_0$, $\| \Xb_0 + \Wb \|_p \leq \| \Xb_0 \|_p$, then with $\Xb_1 = - \Ub \diag (\Sigb(\Xb_0)) \Vb^T$, we have $\| \Xb_1 + \Wb \|_p \leq \| \Xb_1 \|_p$.
\begin{IEEEproof}
The proof easily follows from \cite[Lemma 2]{OymaMFH11} by replacing $\| \cdot \|_*$ with $\| \cdot \|_p$ and applying inequality \eqref{mainInEq}.
\end{IEEEproof}
\end{Lem5}

\subsection{Proofs} \
\begin{IEEEproof}[Proof of Lemma \ref{NSPNec}]
If $\ACal(\Xb) = \bb$, then all feasible solutions to \eqref{pNM} can be represented as $\Xb + \Wb$ for some $\Wb \in \NullDef$. Consequently, to prove that $\Xb$ is a unique solution to \eqref{pNM}, we need to show that $\|\Xb + \Wb\|_p^p > \|\Xb\|_p^p$ for all $\Wb \in \NullDef$. Applying Lemma \ref{SubAddCor}, it can be written that
\begin{IEEEeqnarray*}{rCl}
%\|\Xb + \Wb\|_p^p & = & \sum_{i=1}^{n} \sigma_i^p(\Xb + \Wb) \geq \sum_{i=1}^{n} |\sigma_i^p(\Xb) - \sigma_i^p(\Wb)| \\
%& = & \sum_{i=1}^{r} |\sigma_i^p(\Xb) - \sigma_i^p(\Wb)| + \sum_{i=r+1}^{n} \sigma_i^p(\Wb) \\
%& \geq & \sum_{i=1}^{r} \sigma_i^p(\Xb) - \sum_{i=1}^{r} \sigma_i^p(\Wb) + \sum_{i=r+1}^{n} \sigma_i^p(\Wb) \\
%& > & \sum_{i=1}^{r} \sigma_i^p(\Xb) = \|\Xb \|_p^p,
\|\Xb + \Wb\|_p^p & = & \sum_{i=1}^{n} \sigma_i^p(\Xb + \Wb) \\
& \geq & \sum_{i=1}^{n} \big\lvert \sigma_i^p(\Xb) - \sigma_i^p(\Wb)\big\rvert \\
& = & \sum_{i=1}^{r} \big\lvert\sigma_i^p(\Xb) - \sigma_i^p(\Wb)\big\rvert + \sum_{i=r+1}^{n} \sigma_i^p(\Wb) \\
& \geq & \sum_{i=1}^{r} \sigma_i^p(\Xb) - \sum_{i=1}^{r} \sigma_i^p(\Wb) + \sum_{i=r+1}^{n} \sigma_i^p(\Wb) \\
& > & \sum_{i=1}^{r} \sigma_i^p(\Xb) = \|\Xb \|_p^p,
\end{IEEEeqnarray*}
which confirms that $\Xb$ is the unique solution.
%Conversely, Let $\Xb' = \Xb + \Wb_{(r)}$ for some $\Wb \in \nullS(\ACal)$, where . We have
%\begin{IEEEeqnarray*}{rCl}
%\|\Xb'\|_p^p = \|\Xb + \Wb_{(r)}\|_p^p = \sum_{i=r+1}^{n} \
%\end{IEEEeqnarray*}
\end{IEEEproof}

\begin{IEEEproof}[Proof of Theorem \ref{GentoMat}]
The proof is a direct consequence of integrating Lemma \ref{InEqLem} of this paper and Theorem 1 and Lemma 5 of \cite{OymaMFH11}.
\end{IEEEproof}

\begin{IEEEproof}[Proof of Theorem \ref{RIPThm2}]
For the sake of simplicity, we prove this theorem for the vector case and by virtue of Theorem \ref{GentoMat} matrix case will follow. Let $\xbs$ denote a solution to \eqref{lpmin_Noisy} and $\vb = \xbs - \xb_0$, where $\xb_0$ is the arbitrary vector we want to recover. Furthermore, let $S_0 \subset \{1,\cdots,n_v\}$ with $|S_0| \leq r$. We partition $S_0^{c} = \{1,\cdots,n_v\}\backslash S_0$ to $S_1,S_2,\cdots$ with $|S_i| = t$ probably except for the last set. As a result, $\vb_{S_i}, i \geq 0$ denote a vector obtained by keeping entries of $\vb$ indexed by $S_i$ and setting all other elements to 0.

Our proof is the same as in \cite[Theorem 3.1]{FoucL09} except the way in which $\|\vb_{S_0}\|_2$ and $\|\vb_{S_1}\|_2$ are bounded. Hence, we use the same notation and only focus on the bounding and omit other details. By applying the RIP definition, we get
\begin{IEEEeqnarray}{rCl}
\|\vb_{S_0} + \vb_{S_1}\|_2^2 & \leq & \frac{1}{1-\dtt} \|\Ab(\vb_{S_0}+\vb_{S_1})\|_2^2 \nonumber \\
&=& \frac{1}{1-\dtt} \langle \Ab(\vb - \sumi \vb_{S_i}),\Ab(\vb - \sumi \vb_{S_i}) \rangle \nonumber \\
&=& \frac{1}{1-\dtt} \Big[\|\Ab \vb\|_2^2 \nonumber + 2 \sumi \langle \Ab \vb,- \Ab \vb_{S_i} \rangle\\
& & \quad  + \sum_{i,j \geq 2} \langle \Ab \vb_{S_i}, \Ab \vb_{S_j} \rangle \Big]. \label{largeInEq2}
\end{IEEEeqnarray}

Now, we find upper bounds for the terms in \eqref{largeInEq2}. %For the first term, it can be observed that
%\begin{equation} \label{InEq21}
%\langle \Ab \vb,\Ab \vb \rangle  \leq \|\Ab \vb\|_2^2.
%\end{equation}
Considering the second term in \eqref{largeInEq2}, it can be written that
\begin{equation} \label{InEq22}
\langle \Ab \vb,-\Ab \vb_{S_i} \rangle \leq \sqrt{1 + \dtt} \|\Ab \vb \|_2 \|\vb_{S_i}\|_2.
\end{equation}
Since $\langle \vb_{S_i}, \vb_{S_j} \rangle = 0$ for $i \neq j$, \cite[Lemma 2.1]{Cand08} implies that
\begin{equation} \label{InEq32}
\langle \Ab \vb_{S_i},\Ab \vb_{S_j} \rangle  \leq  \dtt \|\vb_{S_i}\|_2 \|\vb_{S_j}\|_2, \quad \forall i \neq j.
\end{equation}
Also,
\begin{equation} \label{InEq42}
\langle \Ab \vb_{S_i},\Ab \vb_{S_i} \rangle  \leq  (1 + \dtt) \|\vb_{S_i}\|_2^2.
\end{equation}
%Putting \eqref{InEq22}, \eqref{InEq32}, and \eqref{InEq42} in \eqref{largeInEq2} and letting $s = \sumi \|\vb_i\|_F$, we get
Putting \eqref{InEq22}-\eqref{InEq42} in \eqref{largeInEq2} and letting $\Sigma = \sumi \|\vb_{S_i}\|_2$, we get
\begin{IEEEeqnarray}{rCl}
\IEEEeqnarraymulticol{3}{l}{\|\vb_{S_0}\|_2^2 + \|\vb_{S_1}\|_2^2} \nonumber \\
& \leq & \frac{1}{1-\dtt} \Big[ \|\Ab \vb\|_2^2 + 2 \sqrt{1 + \dtt} \|\Ab \vb\|_2 \Sigma \nonumber \\
&& + \dtt \sum_{\substack{i,j \geq 2\\ i \neq j}} \|\vb_{S_i}\|_2 \|\vb_{S_j}\|_2 + (1 + \dtt) \sumi \|\vb_{S_i}\|_2^2 \Big] \nonumber \\
& = & \frac{1}{1-\dtt} \Big[ \|\Ab \vb\|_2^2 + 2 \sqrt{1 + \dtt} \|\Ab \vb\|_2 \Sigma + \dtt \Sigma^2 \nonumber \\
&&  + \sumi \|\vb_{S_i}\|_2^2 \Big] \nonumber \\
& \leq & \frac{1}{1-\dtt} \Big[ \|\Ab \vb\|_2^2 + 2 \sqrt{1 + \dtt} \|\Ab \vb\|_2 \Sigma \nonumber \\
&& + (\dtt + 1) \Sigma^2 \Big] \label{InEq52}
\end{IEEEeqnarray}
where, for the last inequality, we use $\sumi \|\vb_{S_i}\|_2^2 \leq \big( \sumi \|\vb_{S_i}\|_2 \big)^2$. Inequality \eqref{InEq52} can be reduced to
\begin{IEEEeqnarray*}{rCl} %\label{InEq63}
\|\vb_{S_0}\|_2 & \leq &  \frac{1}{\sqrt{1-\dtt}} \big[ \|\Ab \vb\|_2 + \sqrt{1 + \dtt} \Sigma \big],\\
\|\vb_{S_1}\|_2 & \leq & \frac{1}{\sqrt{1-\dtt}} \big[ \|\Ab \vb\|_2 + \sqrt{1 + \dtt} \Sigma \big].
\end{IEEEeqnarray*}
%where $s = \sumi \|\vb_i\|_F$.

The rest of the proof is similar to \cite[Theorem 3.1]{FoucL09} with new parameters $\lambda = 2/\sqrt{1- \dtt}$ and $\mu = \sqrt{1 + \dtt}/\sqrt{1 - \dtt} (r/t)^{\frac{1}{p}-\frac{1}{2}}$. Therefore, in this proof, from $\mu < 1$, we get
\begin{equation*}
\dtt < \frac{(t/r)^{\frac{2}{p}-1} - 1}{(t/r)^{\frac{2}{p}-1} + 1},
\end{equation*}
and, after some simple algebraic manipulations, we obtain
\begin{IEEEeqnarray*}{rCl}
& \|\xb_0-\xbs\|_p \leq &  C_1'  \|\xb_0 - \xb_0^{(r)} \|_p + D_1' r^{\frac{1}{p}-\frac{1}{2}} \epsilon, \\
& \|\xb_0-\xbs\|_2 \leq &  C_2' t^{\frac{1}{2} - \frac{1}{p}} \|\xb_0 - \xb_0^{(r)} \|_p + D_2'  \epsilon,
\end{IEEEeqnarray*}
with constants
\begin{align*}
C_1' &= \frac{2^{\frac{2}{p}-1} (1+\mu^p)^{\frac{1}{p}}}{(1-\mu^p)^{\frac{1}{p}}}, \quad D_1' = \frac{2^{\frac{2}{p}-1} \lambda}{(1-\mu^p)^{\frac{1}{p}}}, \\
C_2' &= \Big( 1 + 2 \sqrt{ \frac{1+\dtt}{1-\dtt} }\Big)\frac{2^{\frac{2}{p}-1}}{(1-\mu^p)^{\frac{1}{p}}}, \\
D_2' &= 2 \lambda + \Big( 1 + 2 \sqrt{ \frac{1+\dtt}{1-\dtt} }\Big) \frac{2^{\frac{1}{p}-1}\lambda}{(1-\mu^p)^{\frac{1}{p}}}.
\end{align*}

\end{IEEEproof}

%\vspace{-49pt}
\section{Conclusion}
In the affine rank minimization problem, it is experimentally verified that Schatten-$p$ quasi-norm minimization is superior to nuclear norm minimization. In this paper, we established a theoretical background for this observation and proved that, under {\color{\MinEdt}a} weaker sufficient condition than that of nuclear norm minimization, global minimization of the Schatten-$p$ quasi-norm subject to compressed affine measurements leads to unique recovery of low-rank matrices. To show that this approach is robust to noise and being approximately low-rank, we generalized some-RIP based results in $\ell_p$ quasi-norm minimization to Schatten-$p$ quasi-norm minimization.

\bibliographystyle{elsarticle-num}
\bibliography{SepSrc}



%\begin{thebibliography}{00}


% \bibitem{cr1}

%\end{thebibliography}

\newpage

\end{document}